# Compact 780-nm Rb Optical Clock

Zhendong Chen[1†], Tianyu Liu[1†], Qiaohui Yang[1], Ya Wang[1], Jie Miao[1], Jingming Chen[1], Ruoao Yang[1],Duo Pan[1*], Shiying Cao[2], Zhigang Zhang[1], Jianjun Wu[1*], and Jingbiao Chen[1,3]

[1] *State Key Laboratory of Advanced Optical Communication Systems and Networks, School of Electronics, Peking University, Beijing 100871, China*
[2]*Division of Time and Frequency Metrology, National Institute of Metrology, Beijing 100013, China*
[3]*Hefei National Laboratory, Hefei 230088, China*
[†]*These authors contributed equally to this work*
** Corresponding author:panduo@pku.edu.cn, just@pku.edu.cn*

ABSTRACT

We demonstrated a compact 780 nm rubidium optical clock, which includes an optical frequency standard and an optical frequency comb, with an optical volume of 11. 6 liters. Unlike the 778 nm rubidium atomic clocks based on two-photon transition, here, the laser frequency is stabilized to the hyperfine transitions of the $^{87}$Rb $5S_{1/2}$ F=2→$5P_{3/2}$ F'=3 , using modulation transfer spectroscopy (MTS). This approach effectively eliminates Doppler background and provides a high signal to noise ratio and high sensitivity. A nearly 300 MHz microwave signal, whose phase exactly tracks that of the optical frequency standard, is generated via the optical frequency comb, yielding a frequency instability of 1.91 E-13 @1 s and 5.29 E-14 @1000 s in the electronic domain. To the best of our knowledge, this is the most precise frequency stabilization result for the first-excited-state transition of alkali metal atoms to date and represents the first optical clock based on this transition. These results offer a promising approach for the development of portable optical clocks.

## 1. Introduction

Optical clocks[1,2], with optical frequency standards and optical frequency combs as fundamental components,[3] have found wide-ranging applications in precision metrology, global time synchronization, and scientific research, particularly in fields such as geodesy,[2,4] quantum information,[5] and fundamental physics.[6] The rapid development of optical frequency standards has accelerated the progress of optical clocks designed for practical applications[7,8]. To replace conventional microwave clocks in practical applications, optical clocks require microwave frequency outputs, which typically necessitate the use of optical frequency combs to down-convert optical frequencies to microwave frequencies. In 2011, the National Institute of Standards and Technology (NIST) realized an optical clock based on Yb atoms, achieving a frequency instability in the microwave range better than $10^{-18}$, representing a two-order-of-magnitude improvement by over the most precise microwave clocks,[9] demonstrating the potential for optical clocks to redefine the standards for frequency measurements in diverse technological applications.

.Optical clocks can be categorized into two primary types based on their intended application: high-precision transportable optical clocks, which prioritize accuracy, and portable optical clocks, which emphasize mobility and ease of deployment.[10-12] To conduct cutting-edge research in fundamental physics, including gravitational red-shift, gravitational wave detection, and dark matter exploration, it is crucial to integrate and transport these complex high-precision optical clocks to various application sites.[13,15] Recent advances in transportable optical clocks have leveraged atomic systems such as neutral atom optical lattices and single ion systems. In 2017, the German Federal Institute of Physics and Technology (PTB) achieved the first portable optical lattice clock for long-distance vehicle transportation and measurement applications. This system, with dimensions of 2.2 m×3 m×2.2 m, demonstrated a stability of $1.3 \times 10^{-15}/\sqrt{\tau}$, and an uncertainty of $7.41\times10^{-17}$.[13] In 2020, the RIkagaku KENkyusho (RIKEN) developed a high-precision transportable optical lattice clock, which was deployed atop the 450-meter Tokyo Skytree to compare with a ground-based optical clock. This work highlighted the potential of optical clocks for field applications, such as crustal deformation and geoid testing, achieving a resolution at the centimeter level.[14]

The compact portable optical clock, leveraging the advantages of optical clocks over microwave clocks in terms of frequency stability, and featuring a small size and low power consumption, is ultimately expected to replace the existing commercial microwave clocks.[17,18] Recent research has mainly focused on the development of thermal atomic/molecular gas-cell optical clocks, which primarily include optical clocks based on iodine molecules and alkali metal atomic gas cells. In 2024, NIST's iodine optical clocks operated continuously aboard a naval ship in the Pacific Ocean for 20 days, with timing errors accumulating to less than 300 picoseconds per day. Their performance is comparable to that of active hydrogen masers, while their volume is 35 L.[18] Optical clocks based on alkali metal atoms, due to the compactness of their optical frequency standards,[19-27] have greater potential for miniaturization and portability. In 2019, The NIST proposed the idea of building a portable optical clock using silicon-based components. They constructed a rubidium two-photon optical clock using wave guide structures, miniature atomic vapor cells, and micro-cavity frequency combs, achieving a microwave signal stability of $4.4\times10^{-12}/\sqrt{\tau}$.[12]

Currently, portable optical clocks predominantly utilize Rb two-photon transition optical frequency standards.[12] However, to date, a optical clock down-conversion for microwave frequencies basd on Rb/Cs first-excited-state transition has not been realized, despite its potential advantages in terms of easier implementation and integration. Additionally, high-precision outputs can be achieved using techniques such as modulation transfer spectroscopy (MTS)[29-33]. In this work, we present the first 780 nm optical clock based on the Rb D2 line. Unlike the two-photon optical clock scheme, the optical frequency standard is locked to the 780 nm $5S_{1/2}$ F=2→$5P_{3/2}$ F'=3 cycling transition of $^{87}$Rb using MTS. To achieve miniaturization of the optical clock, a custom-designed optical frequency comb based on femtosecond fiber lasers has been developed. The in-loop frequency instability of optical frequency comb exceeds the $10^{-17}$ level, with a carrier-envelope offset (CEO) frequency ($f_{ceo}$) stability of $7.86\times10^{-18}$ @1s and a beat signal between the optical frequency comb and optical frequency standard ($f_{b780\ nm}$) of $1.01\times10^{-17}$ @1s, normalized to the optical frequency. The integrated optical frequency comb effectively transfers the clocks' optical frequency stability into its microwave outputs, with stability reaching $1.91\times10^{-13}$ at 1-s and $5.29\times10^{-14}$ at 1000-s. The short-term stability is limited by the hydrogen maser reference. Among the reported frequency stabilization results for the D1 and D2 lines of alkali metals atoms, our results are competitive, and the optical clock system has been designed in an integrated manner, featuring portability and facilitating ease of application. These results demonstrate the successful miniaturization and extraordinary stability of a 780 nm optical clock, offering new possibilities for portable and high-precision optical clocks in applications such as global navigation satellite systems (GNSS) and quantum metrology.

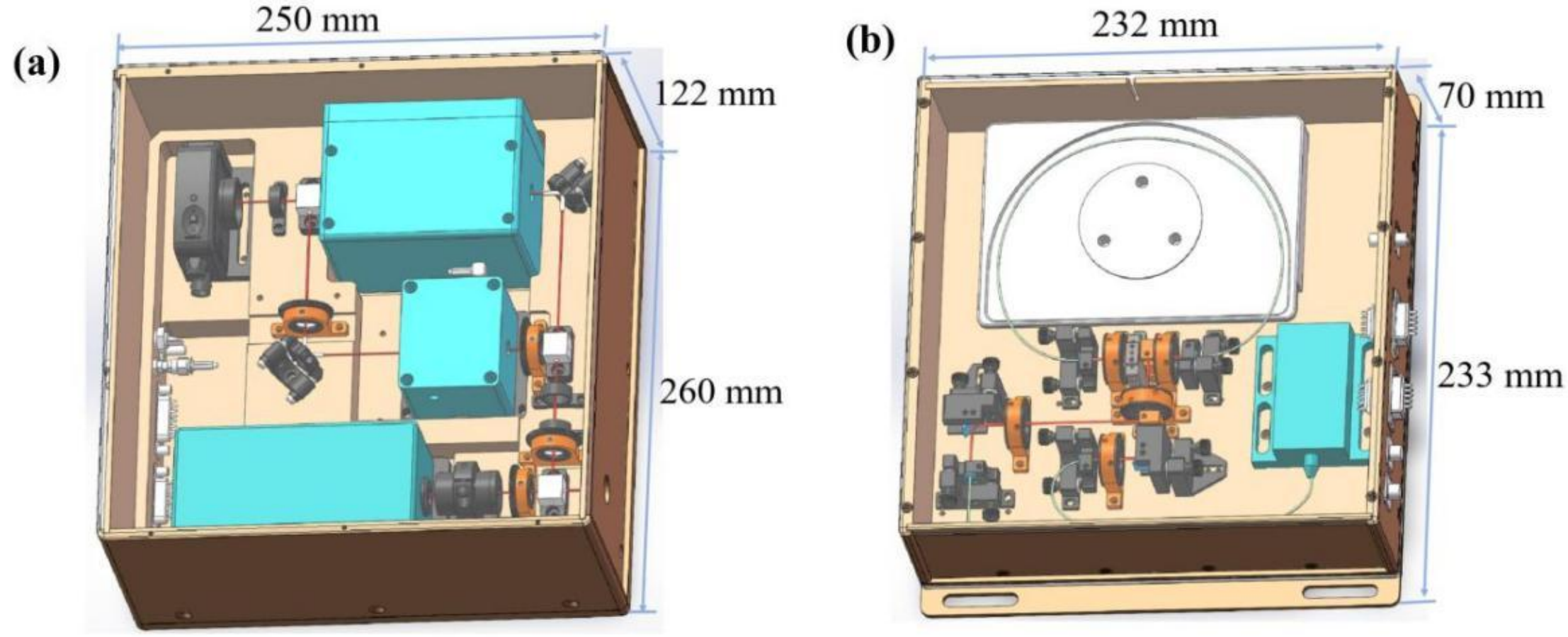

Fig. 1. (a) Experimental setup of 780 nm optical frequency standard based on MTS system. (b) Experimental setup of optical frequency comb based Erbium-doped fiber laser.

## 2. Experimental methods

Fig. 1 shows the experimental setup for both the optical frequency standard and the optical frequency comb. To enhance mechanical stability, all components are firmly attached to an aluminum baseboard, and the entire setup is enclosed during operation to minimize the impact of external disturbances.

The schematic of the Rb 780-nm optical frequency standard based on MTS is shown in Fig. 2(a). A homemade 780-nm interference-filter external cavity diode laser (ECDL) serve as the local oscillator. The laser frequency is controlled by adjusting the injection current and the piezoelectric transducer (PZT). The laser, after traversing an isolator and a beam expander, emits a beam at 780.246 nm, which is split by a PBS into two beams: one for the pump and the other for the probe. The power ratio between the pump and probe beams can be adjusted by rotating the two wave plates along the optical path. By rotating the wave plate, the polarization direction of the pump beam is aligned with the axis of the electro-optic modulator (EOM). To minimize residual amplitude modulation of the EOM, an EOM with beveled edges is used, and the temperature of the EOM is controlled by a temperature controller. The pump light undergoes phase modulation via the EOM before entering a φ10 mm × 50 mm cylindrical glass cell filled with $^{87}$Rb atoms. The probe light enters the gas cell and overlaps with the pump light in the opposite direction. The gas cell is thermally insulated with multiple layers of Teflon and enclosed within a double-layer cylindrical permalloy shield for magnetic protection. Simultaneously, the temperature of the rubidium atomic vapor cell is monitored by a thermistor, and regulated by a temperature controller, which adjusts the current to the heater accordingly. The MTS optical system is enclosed in a metal box, which is wrapped in teflon, with a total volume of 7.8 L.

The modulation transfer process takes place within the rubidium atomic vapor cell. The probe light is detected by the photodetector, the signal is then demodulated by a mixer to obtain the MTS signal, which serves as the error signal for the proportional-integral-derivative (PID) locking system. The rubidium atom, selected as the quantum reference (the relevant energy levels are shown in Fig. 2, allows the 780 nm laser to be locked to the $^{87}$Rb fine level $5S_{1/2}$ F=2→$5P_{3/2}$ F'=3 clock transition by the driving current and the PZT.

The experimental setup of the entire optical clock system is shown in Fig. 3, where the optical path is represented by red lines, and the electrical circuit is represented by gray lines. The femtosecond laser source is a home-made Er:fiber laser based on nonlinear polarization evolution (NPE) mode-locking mechanisms. The laser provides an output power of 280 mW, with a fundamental repetition rate of up to 282 MHz. The net dispersion of the laser cavity is designed to be near zero, facilitating stretched-pulse operation and thereby reducing noise levels. An EOM and a PZT, attached to the fiber as feedback devices, are incorporated into the laser system. The oscillator is placed in an aluminum box wrapped in thermal insulation, with thermoelectric coolers (TECs) located underneath the inner aluminum box.

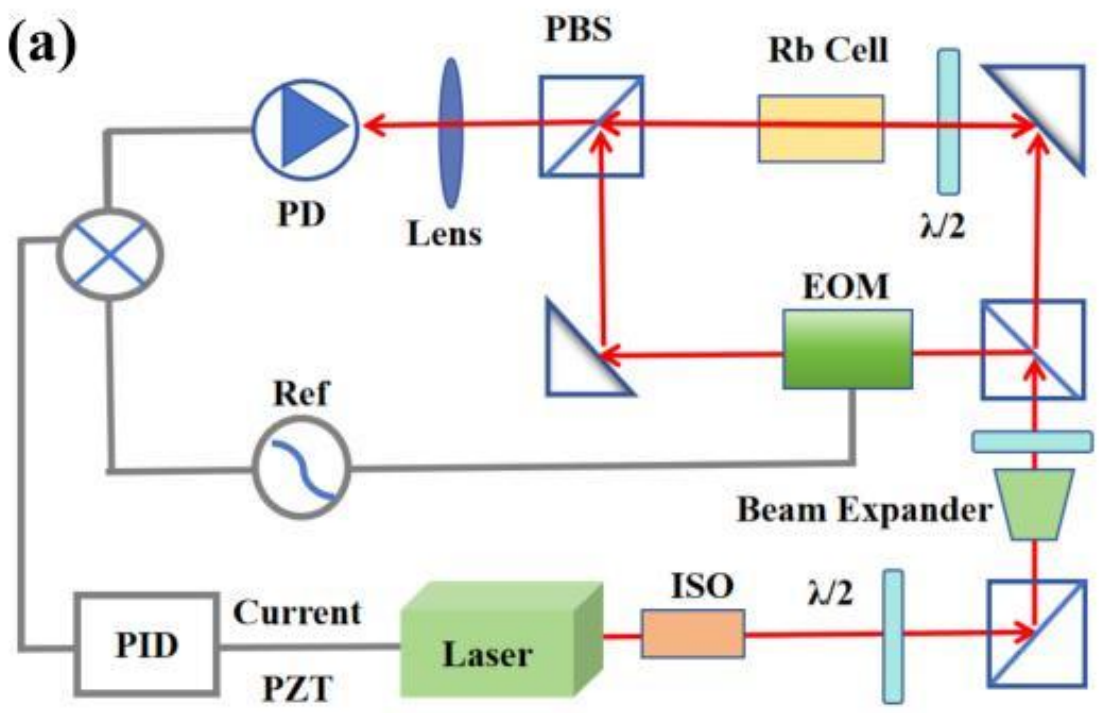

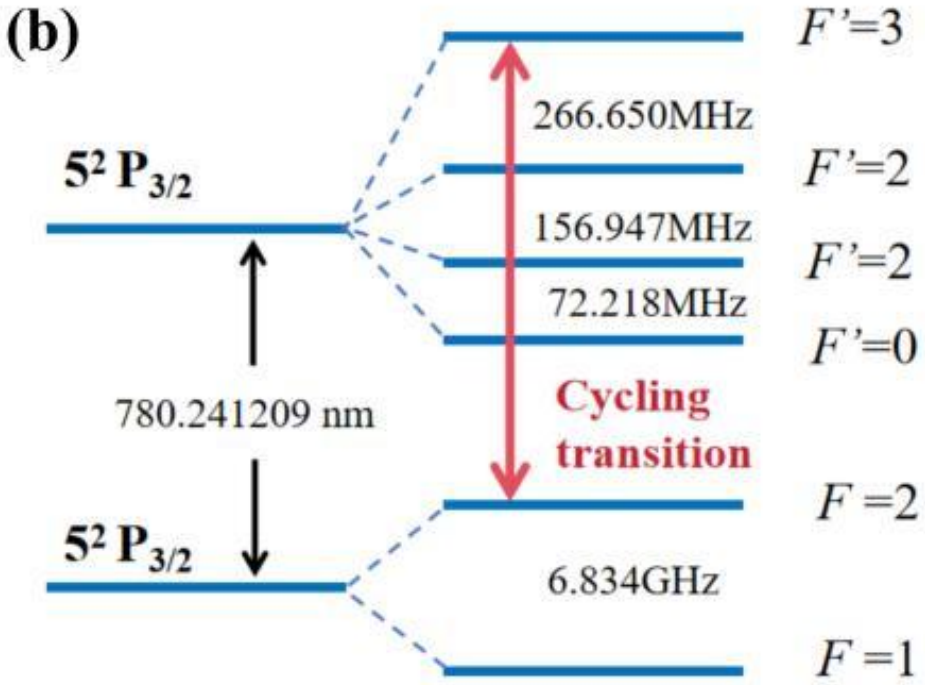

Fig. 2. (a) Schematic diagram of the 780 nm frequency standard based on MTS system: λ/2 is a half-wave plate, PZT is a piezoelectric transducer, ISO is a space isolator, PBS is a polarizing beam splitter, EOM is an electro-optic modulator, PD is a photodetector. (b) hyperfine levels of the $^{87}$ Rb D2 transition

The laser output from the Erbium-doped fiber laser is split into two branches by the PBS. One beam is used to generate the arrier-envelope offset signal ($f_{ceo}$), while the other beam is directed to beat with the optical frequency standard. To facilitate spectral broadening, an erbium-doped fiber amplifier (EDFA) is constructed. After optimization, the amplifier produces an output power of 120 mW with a pulse width of 69 fs, which are coupled to a f-to-2f waveguide module based on a tantalum pentoxide(Octave Photonics, COSMO),). In this waveguide module, the spectrum is broadened to an octave-spanning range extending from 1100 nm to 2200 nm, and the $f_{ceo}$ is obtained by doubling the frequency. The $f_{ceo}$ signal is first filtered through a band-pass filter (BPF) to reduce background noise, then amplified to around 0 dBm before entering the phase-locking circuit. The pump current of the oscillator is controlled by the locking circuit to stabilize the fceo signal, which is referenced to the rubidium clock. The second beam of the Erbium-doped fiber laser, with a power of 20 mW, is amplified to 150 mW by another EDFA. The resulting 1560 nm laser is then frequency doubled through a periodically-poled lithium niobate doped with magnesium oxide (MgO: PPLN) crystal, with a poling period of 19.60 μm. To optimize frequency doubling efficiency, focusing and collimating lens are placed at the front and back of the crystal, respectively. This setup yields an output power of 8 mW at 780 nm (with a bandwidth of 10 nm). The optical frequency standard, which outputs a 780 nm continuous-wave laser, is collimated and aligned with the optical frequency comb laser beam using a lens to match the spot sizes. The two beams are combined using a PBS, with wave plates adjusted to generate a beat frequency signal ($f_{b780nm}$). The beat signal between the optical frequency comb and the optical frequency standard is locked to the commercial rubidium clocks using a phase-locked loop circuit. A PZT functions as the slow loop, while an EOM provides fast loop fine-tuning. As a result, each tooth of the optical frequency comb is stabilized to the 780 nm Rb optical frequency standard, forming an integrated optical clock system. Therefore, the stability of the optical frequency standard can be down-converted to microwave output through the optical frequency comb.

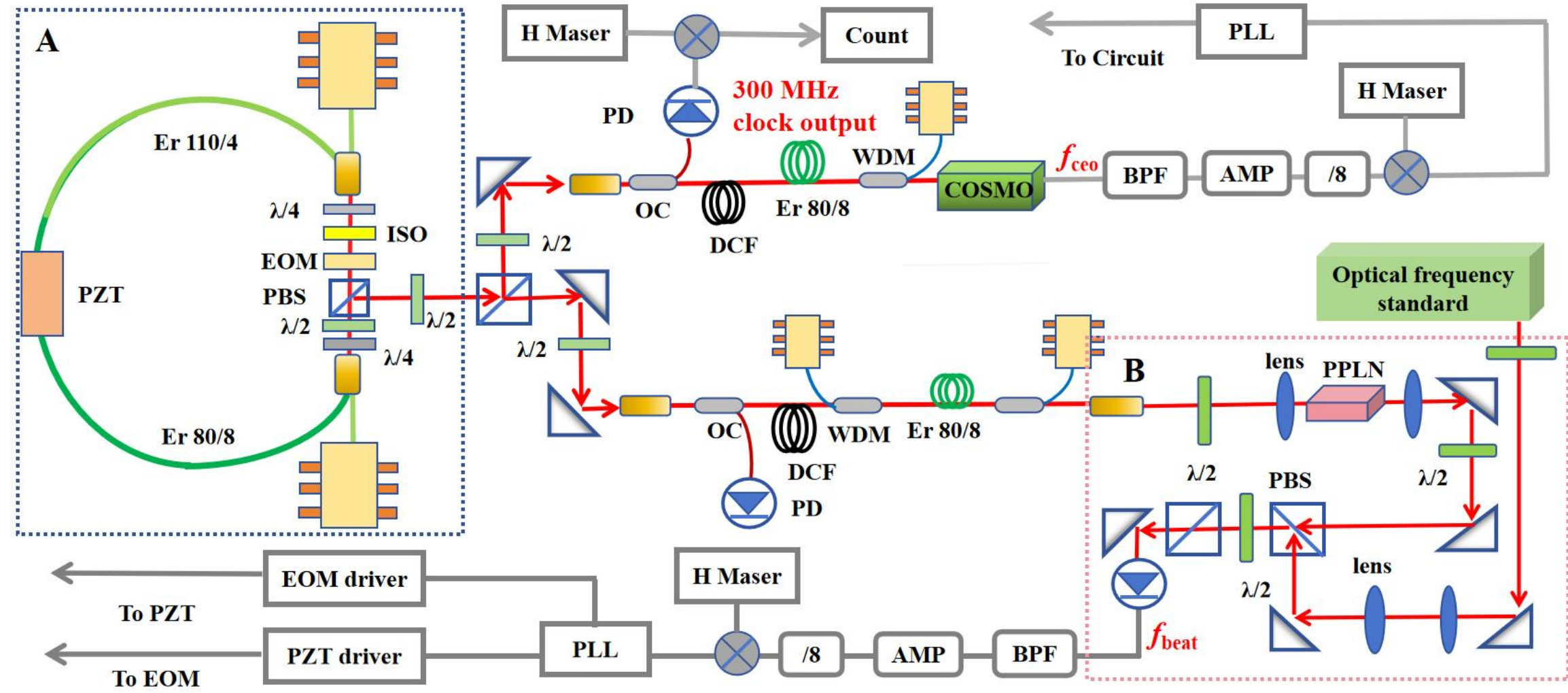

Fig. 3. Structure of compact 780 nm Rb optical clock. ISO is a space isolator, λ/4 is a quarter-wave plate, λ/2 is a half-wave plate, PZT is a piezoelectric transducer, WDM is wavelength division multiplexing component, PBS is a polarizing beam splitter, PC is a polarization controller, EOM is an electro-optic modulator, PD is a photodetector, PPLN is a frequeny conversion crystal, H-maser is a microwave reference, AMP is a microwave amplifier, BPF is a band-pass filter, and PLL is a phase lock loop. Part A is an femtosecond laser, Part B is beat frequency module.

## 3. Results and discussion

### 3.1. 780nm Rb optical frequency standards

The MTS lineshape varies with the gas cell temperature and modulation frequency, with the slope at the center point of the demodulation signal reflecting the spectral line's sensitivity. At a modulation frequency of 9.62 MHz and a gas cell temperature of 37.08 °C, the MTS achieves its maximum slope of 542 mV/MHz. The quantum reference spectrum of $^{87}$Rb $5S_{1/2}$ F=2→$5P_{3/2}$ F'=3 is depicted in Fig. 4(a), where the blue curve represents the Saturated absorption spectrum (SAS) signal detected by the photodetector, and the red curve represents the MTS signal. Using the frequency discrimination signal from the MTS, we lock the 780 nm continuous-wave laser to the D2 line ($^{87}$Rb $5S_{1/2}$ F=2→$5P_{3/2}$ F'=3) of $^{87}$Rb through optimizing the parameters of the Proportion Integration Differentiation (PID) circuit. The true value of the optical frequency standard is measured using an optical frequency comb whose initial frequency and repetition rate are fully locked to the hydrogen maser clock. Fig. 4. (b) displays the beat frequency count between the optical frequency comb and optical frequency standard. The true value of the optical frequency standard, determined based on the optical frequency comb's repetition frequency and initial frequency, is 384,228,115,588.473 kHz.

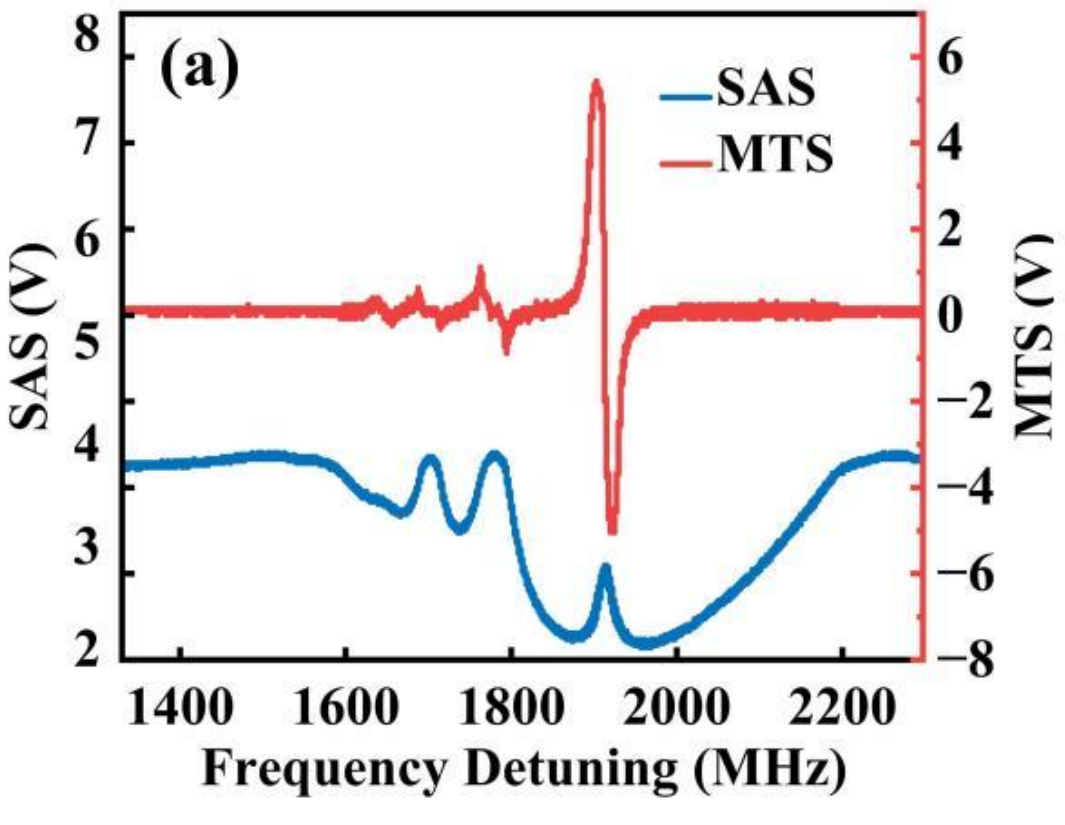

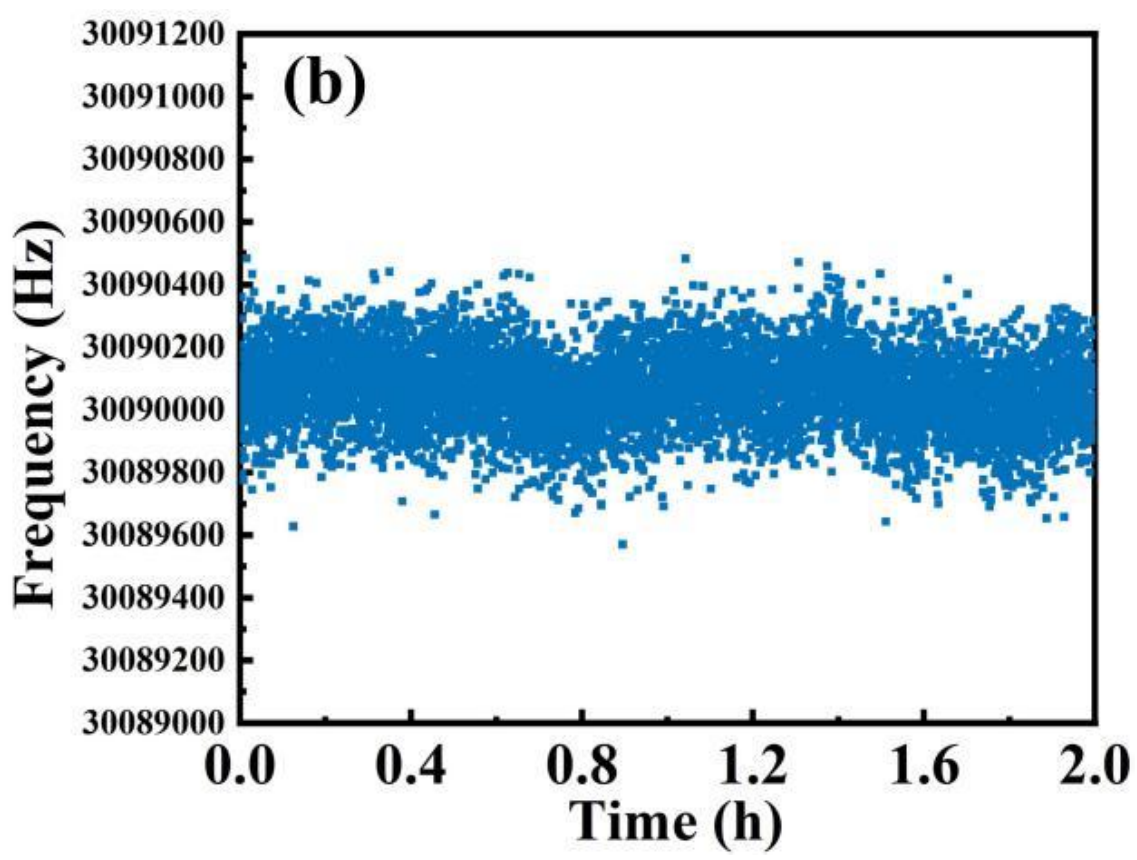

Fig. 4. (a) Spectra signals of $^{87}$Rb $5S_{1/2}$ F=2→$5P_{3/2}$ F'=3 modulation transfer spectrum (red line) and saturated absorption spectroscopy（SAS, blue line). (b) the fluctuations of the $f_{beat}$ signal between the optical frequency comb and optical frequency standard.

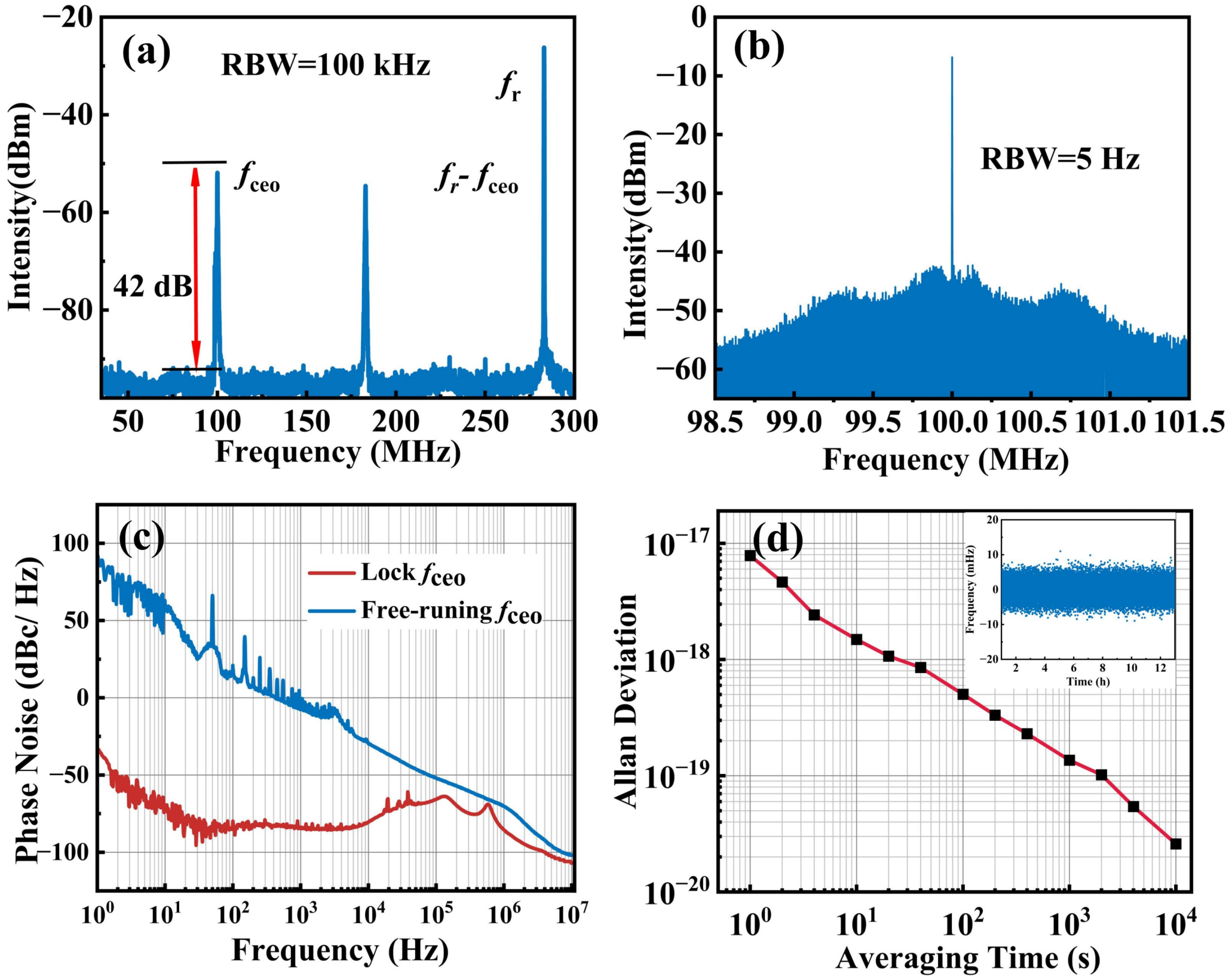

Fig. 5. (a) The frequency of free-runing $f_{ceo}$. (b) The frequency of phase-locked $f_{ceo}$. (c) The power spectral density of free-runing $f_{ceo}$ (blue) and phase-locked $f_{ceo}$ (rad). (d) The in-loop relative frequency instability of phase-locked $f_{ceo}$ (it is normalized to the optical frequency). Inset: the fluctuations of the phase-locked $f_{ceo}$ signal for about 12 hours.

### 3.2. optical frequency comb

To convert the optical frequency of the rubidium atomic frequency standard to microwaves, the initial frequency of the home-made optical frequency comb must first be locked to the hydrogen clock. As shown in Fig. 5 (a), the signal-to-noise ratio of the free-running $f_{ceo}$ is estimated to be 41 dB at a resolution bandwidth of 100 kHz. The $f_{ceo}$ signal is locked to 100 MHz by optimizing the parameters of the locking circuit. From the servo peak, the servo bandwidth is estimated to be 710 kHz, with a signal-to-noise ratio of 37 dB - see Fig. 5 (b). Fig. 5 (c) illustrates the phase noise of the $f_{ceo}$ under free-running and locked conditions. Phase-locking significantly reduces the phase noise, with a substantial decrease from 92 dBc/Hz to -33 dBc/Hz at a 1 Hz frequency offset. The integrated phase noise of the locked $f_{ceo}$ is 388 mrad, integrated from 10 MHz to 1 Hz. Frequency counts of the locked $f_{ceo}$ , recorded over 12 hours, show a standard deviation of 2.33 mHz, as depicted in Fig. 5 (d). To demonstrate the stability of $f_{ceo}$, the Allan deviation is calculated, yielding a frequency stability of $7.86\times10^{-18}$ at 1s, and drops to $2.59\times10^{-20}$ at 10000 s, which is normalized to the optical frequency.

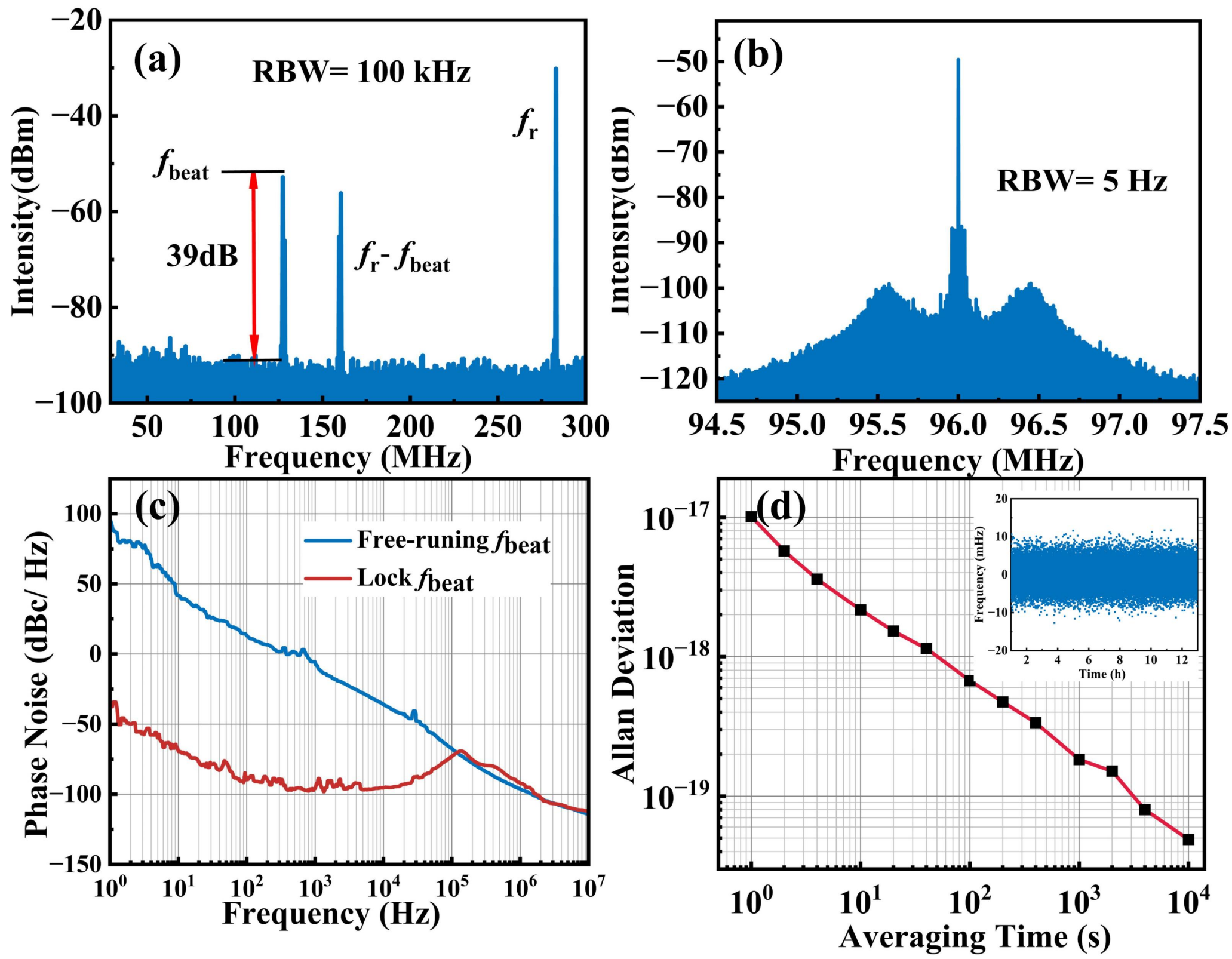

Fig. 6. (a) The frequency of free-runing $f_{b780\ nm}$. (b) The frequency of phase-locked $f_{b780\ nm}$. (c)The power spectral density of free-runing $f_{b780\ nm}$ (blue) and phase-locked $f_{b780\ nm}$ (rad). (d) The in-loop relative frequency instability of the phase-locked $f_{b780\ nm}$. (it is normalized to the optical frequency). Inset: the fluctuations of the phase-locked $f_{b780\ nm}$ signal for about 12 hours.

It is observed that a higher beat frequency signal-to-noise ratio (SNR) facilitates easier locking. By optimally adjusting the spot sizes and the overlap of the two optical beams, the SNR of the free-running $f_{b780\ nm}$ reaches 39 dB, as shown in Fig. 6(a). It is further improved to 55dB after locking, with a bandwidth of 450 kHz, as depicted in Fig. 6(b). Fig. 6(c) illustrates the phase noise of the $f_{b780\ nm}$, which is significantly suppressed through phase-locking. The root mean square (RMS) phase noise of phase-locked $f_{b780\ nm}$ is 358 mrad, integrated over the range from 10 MHz to 1 Hz. The stability of $f_{b780\ nm}$ signal is depicted in Fig. 6 (d), where the Allan deviation is $1.01 \times 10^{-17}$ at 1 s. Data from continuous frequency counting over 12 hours, depicted in the illustration of Fig. 6 (d), yields a standard deviation of 3.04 mHz. Therefore, the optical frequency comb's tracking stability with respect to the optical frequency standard is sufficient to transfer the optical stability of the optical clocks into the microwave domain.

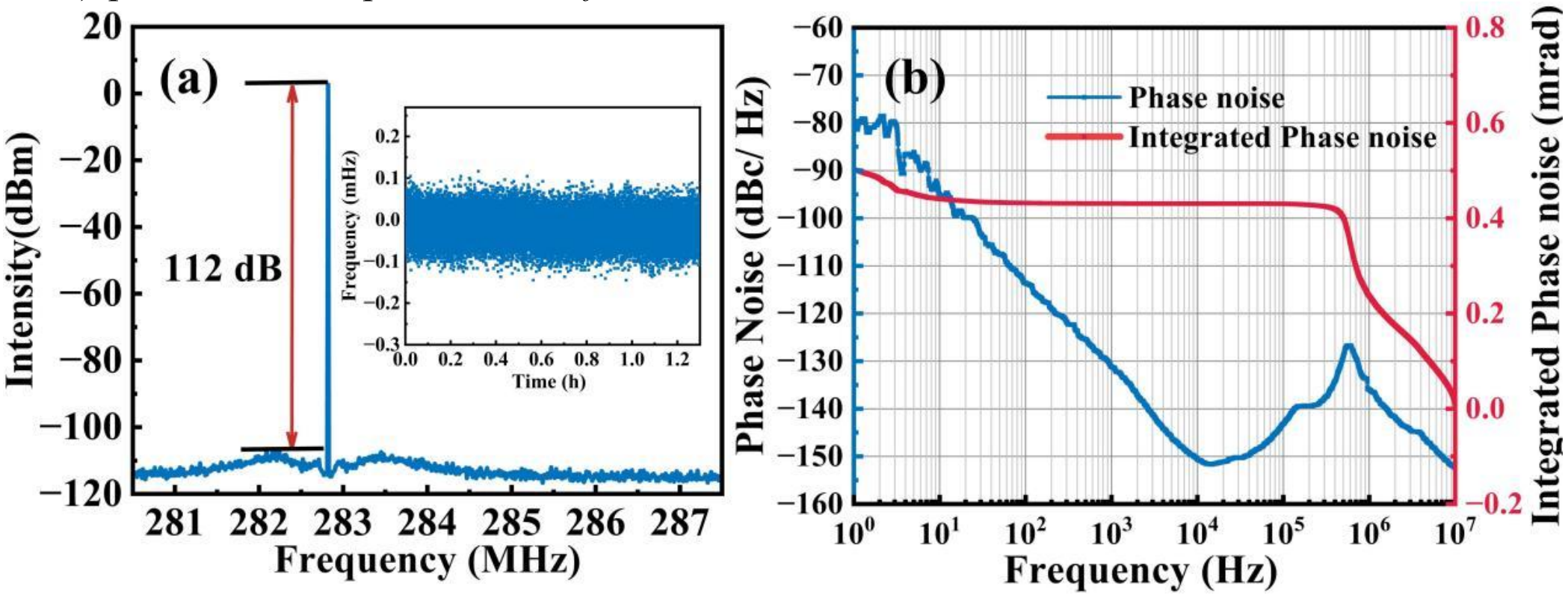

Fig. 7. (a) The performance of optical clock down-conversion to microwave frequencies. Inset: the frequency fluctuation of the optical clock signal. (b) The measured phase noise spectrum of optical clock down-conversion to microwave frequencies (blue line) and an integrated (1 Hz-10 MHz) root mean square phase noise (red line).

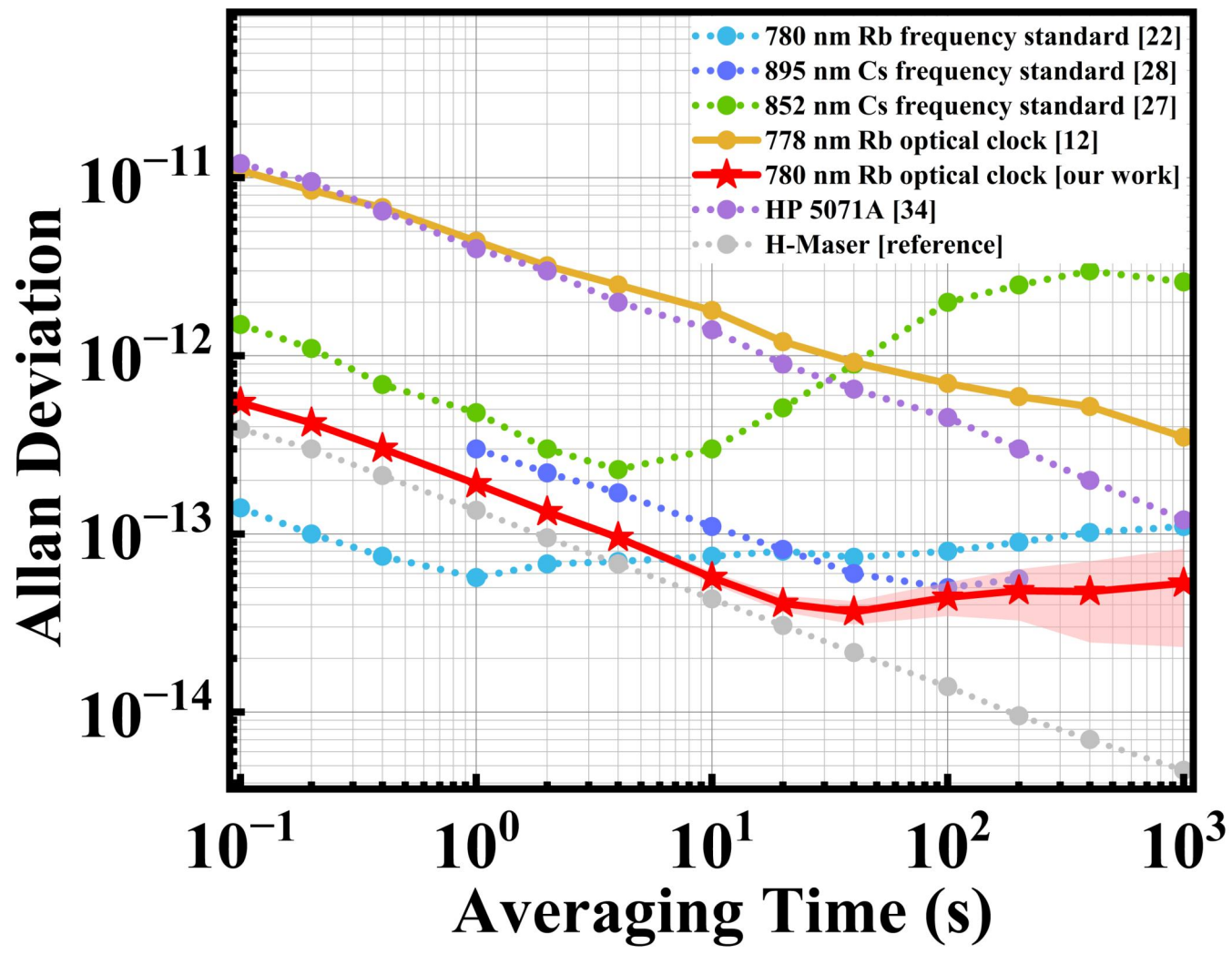

Fig. 8. The frequency stability of the thermal alkali metal atomic frequency standards and optical clocks. (the solid line represents an optical clock, and the dashed line represents an optical frequency standard).

### 3.3. optical clock

As the optical frequency comb is phase locked to the optical frequency standard, the repetition rate of the optical frequency comb in the microwave range inherits the stability of the optical frequency standard. The signal-to-noise ratio (SNR) of the microwave signal, as depicted in Fig. 7(a), is 112 dB at a 3 Hz resolution bandwidth (RBW). The corresponding microwave frequency emitted by the optical clock is illustrated in illustration of Fig. 7(a) with a standard deviation estimated to be 0.033 mHz. The phase noise power spectral density is measured by a phase noise analyzer as shown in Fig. 7(b), which complements the short-term frequency stability measurements.

As shown in Fig. 8, the short-term stability of the compact rubidium 780 nm optical clock is $1.91\times10^{-13}$ at 1 s, reaching $3.65\times10^{-14}$ at 40 s, which is limited by the stability of the reference hydrogen maser. The Allan deviation slightly degrades thereafter, but can still reach $5.29\times10^{-14}$ at 1000 s. This stability is highly competitive with previously reported thermal alkali metal atomic frequency standards and optical clocks. Furthermore, unlike the two-photon system that requires fluorescence collection equipment, this optical clock employing MTS provides a novel and portable solution.

## 4. Conclusion

In conclusion, we report the first compact 780 nm Rb optical clock down-conversion for microwave frequencies. The optical volume of the optical clock has been reduced to 11.6 L by optimizing optical frequency standard's structure and employing a self-made compact optical frequency comb. The integrated optical frequency comb is phase-locked to the optical frequency standard, enabling optical frequency down-conversion to a microwave output. The Allan deviation is calculated to be $1.91\times10^{-13}$ @1s, reaching $3.65\times10^{-14}$ at 40 s, and does not exceed $5.29\times10^{-14}$ at 1000s. To the best of our knowledge, this is the first compact 780 nm Rb optical clock based on the Rb D2 transition. These results provide a promising approach for portable optical clocks and pave the way for their broader applications. Future work will focus on enhancing the long-term stability of the optical clock by more precisely temperature control and velocity-comb modulation transfer spectroscopy.

## ACKNOWLEDGMENT

This work is supported by Beijing Nova Program (No. 20240484696); INNOVATION Program for Quantum Science and Technology (2021ZD0303200); Wenzhou Major Science and Technology Innovation Key Project (No. ZG2020046); National Natural Science Foundation of China (U2031208); Boya Postdoctoral Fellowship and Postdoctoral Fellowship Program (GZB20230009).

## AUTHOR DECLARATIONS

**Conflict of Interest**

The authors have no conflicts to disclose.

**Author Contributions**

Zhendong Chen and Tianyu Liu contributed to the work equally and should be regarded as co-first authors.

**Zhendong Chen**: Formal analysis (equal); Investigation (equal); Methodology (equal); Software (equal); Validation (equal); Writing – original draft (equal); Writing – review & editing (equal). **Tianyu Liu**: Formal analysis (equal); Investigation (equal); Methodology (equal); Software (equal); Validation (equal); Writing – original draft (equal); Writing – review & editing (equal). **Qiaohui Yang**: Formal analysis (equal); Investigation (equal); Methodology (equal). **Jie Miao:** Formal analysis (equal); Software (equal); Writing – original draft (equal); Writing – review & editing (equal). **Jingming Chen**: Formal analysis (equal); Investigation (equal). Ya Wang: Software (equal); Writing – review & editing (equal). **Duo Pan:** Conceptualization (equal); Data curation (equal); Funding acquisition (equal); Project administration (equal); Supervision (equal); Writing – review & editing (equal). **Ruoao Yang:** Conceptualization (equal); Data curation (equal); Investigation (equal); Methodology (equal); Writing – original draft (equal); Writing – review & editing (equal). **Jianjun Wu**: Conceptualization (equal); Data curation (equal); Investigation (equal); Methodology (equal); Writing – original draft (equal); Writing – review & editing (equal). **Shiying Cao:** Conceptualization (equal); Data curation (equal); Investigation (equal); Methodology (equal); Writing – original draft (equal); Writing – review & editing (equal). **Zhigang Zhang:** Conceptualization (equal); Data curation (equal); Investigation (equal); Methodology (equal); Writing – original draft (equal). **Jingbiao Chen:** Conceptualization (equal); Data curation (equal); Funding acquisition (equal); Project administration (equal); Supervision (equal); Writing – review & editing (equal).